\pdfoutput=1

\documentclass[pra,10pt,superscriptaddress, footnoteinbib]{revtex4}
\usepackage{amsmath}
\usepackage{latexsym}
\usepackage{amssymb}
\usepackage{graphics,epstopdf}
\usepackage[colorlinks=true, citecolor=blue, urlcolor=blue ]{hyperref}
\usepackage{epsf,graphics,graphicx}

\newcommand{\n}{\noindent}

\newcommand{\ed}{\end{document}}
\newcommand{\beq}{\begin{equation}}
\newcommand{\eeq}{\end{equation}}

\begin{document}
\title{The geometric phase and the geometrodynamics of relativistic electron vortex beams }
\author{{Pratul Bandyopadhyay \footnote{Electronic address:{b$_{-}$pratul@yahoo.co.in}} ${}^{}$}, Banasri Basu\footnote{Electronic
address:{sribbasu@gmail.com}}${}^{}$, Debashree Chowdhury\footnote{Electronic
address:{debashreephys@gmail.com}}${}^{}$ }
\affiliation{Physics and
Applied Mathematics Unit, Indian Statistical Institute,\\
 203
Barrackpore Trunk Road, Kolkata 700 108, India}


\begin{abstract}
\n
We have studied here the geometrodynamics of relativistic electron vortex beams from the perspective
of the geometric phase associated with the scalar electron encircling the vortex line.
It is pointed out that the electron vortex beam carrying orbital angular momentum is a natural consequence of
the skyrmion model of a fermion. This follows from the quantization procedure of a fermion in the framework of
Nelson's stochastic mechanics when a direction vector (vortex line) is introduced to depict the spin degrees of
freedom. In this formalism a fermion is depicted as a scalar particle encircling a vortex line.
It is here shown that when the Berry phase acquired by the scalar electron encircling the vortex
line involves quantized Dirac monopole we have paraxial (non-paraxial) beam when the vortex line is parallel
(orthogonal) to the wavefront propagation direction. Non-paraxial beams incorporate spin-orbit interaction.
When the vortex line is tilted with respect to the propagation direction the Berry phase involves non-quantized monopole. The temporal variation of the direction of the tilted vortices is studied here taking into account the renormalization group flow of the monopole charge and it is predicted that this gives rise to spin Hall effect.

\end{abstract}

\maketitle
~~~~~~~~~~~~~~~$keywords$:~~~    electron vortex beam, Berry phase, spin Hall effect
\section{Introduction} \label{sec1}

In a seminal paper, Nye and Berry (1974) pointed out that in 3D space the wave fronts in general contain dislocation lines when the phase becomes singular and currents coil around the vortex line. These are known as optical vortices.
The quantized vortex strength corresponds to a topological charge. When the vortex line is parallel (orthogonal) to the wave propagation direction the corresponding situation is characterized as screw (edge) dislocations. It is also possible to have mixed screw-edge dislocations characterized by the vortex lines which are tilted with respect to the propagation direction. Screw dislocations arise for monochromatic waves whereas mixed screw-edge dislocations require temporal variations. Allen et. al. (1992) demonstrated that optical beams in free space bearing screw dislocation
posses quantized orbital angular momentum (OAM) directed along the beam axis. This suggests that for monochromatic vortex beams the intrinsic OAM is collinear  to the momentum and its projection on the beam axis takes quantized values. Bliokh et. al (2012) have shown that it is also possible for a wave front with mixed screw-edge dislocations having tilted vortex lines to carry well defined OAM in an arbitrary direction. These correspond to time diffracting or non-diffracting spatio-temporal vortex beams which are polychromatic in nature.

Recently, electron vortex beams with OAM have been produced experimentally (Uchida and Tonomura 2010, Verbeeck et. al. 2010, McMorran et. al. 2011). This bears a close analogy between optical and matter waves. Electron vortex beams are generally visualized as scalar electrons orbiting around vortex lines. Indeed, Bliokh et. al. (2007) earlier predicted the existence of free space vortex beams for nonrelativistic scalar electrons. Bliokh et. al. (2011) studied the relativistic electron vortex beams representing the angular momentum eigenstates of a free Dirac electron and constructed exact Bessel beam solutions.  These authors considering the spin-orbit interaction (SOI) gave a self  consistent description of the OAM and spin angular momentum (SAM) properties of the Dirac electron. Bessel beams in general represent monoenergetic plane waves having constant momentum generating a fixed polar angle with the z-axis.
In the limit of vanishing SOI the solutions are eigenstates of both OAM and SAM. This occurs for the polar angle $\theta=0$ when we have paraxial beams and also in the non-relativistic case implying momentum $p\rightarrow 0$. However, when we switch on SOI, in general, we have non-paraxial beams which are eigenvalues of the total angular momentum but not of the OAM and  SAM separately.

In this note we study the geometrodynamics of relativistic electron vortex beams from the perspective of the geometric phase (Berry 1984) acquired by the scalar electron encircling the vortex line. It has been pointed out earlier (Bandyopadhyay and Hajra 1987, Hajra and Bandyopadhyay 1991) that the quantization of a fermion in the framework of Nelson's stochastic quantization procedure (Nelson 1966, 1967) can be achieved when we introduce a direction vector at the space-time point which appears  as a vortex line depicting the spin degree of freedom. In this scenario, a fermion is depicted  as a scalar particle encircling a vortex line which is topologically equivalent to a magnetic flux line. This effectively gives rise to a gauge theoretic extension of the space-time coordinate as well as momentum and the spin appears as an SU(2) gauge bundle. In this framework, a massive fermion appears as a skyrmion (Skyrme 1961, 1962). The specific fermionic properties such as the sign change of the wave function after a $2\pi$ rotation as well as the spin-statistics relation is a manifestation of the Berry phase acquired by the scalar electron orbiting around the vortex line (Bandyopadhyay 2010). When the scalar particle rotating around the vortex line acquires certain quantized OAM $l,$ the total phase acquired by it is $2\pi l+\phi_B,$ $\phi_B$ being the Berry phase. For a quantized OAM, with $l\in Z,$  the effective phase for such a system is $\phi_B$ as the factor $2\pi l$ leads to a trivial phase. So when the system bears certain quantized OAM the specific fermionic properties generated through the Berry phase are not disturbed. In this framework, a relativistic electron vortex beam carrying quantized OAM appears to be a natural consequence. The geometrodynamics of such beams correspond to situations when vortex lines are oriented along different directions. In fact, the Bessel beam spectrum forms a cone of plane waves with a fixed polar angle $\theta$ with the quantization axis (z-axis). When the vortex line is along the z-axis ($\theta=0$) which corresponds to the wave front propagation direction, we have paraxial beams. Non-paraxial beams are generated when we switch on SOI. The Berry phase acquired by the scalar electron orbiting around the vortex line involves quantized Dirac monopole when the polar angle $\theta$ with the z-axis is zero and $\pi/2$. But for other angular orientations of the vortex line, the concerned monopole is non-quatized. It has been pointed out that the monopole charge in 3+1 dimension is equivalent to the central charge of the conformal field theory in 1+1 dimensions (Bandyopadhyay 2000,2011). As the central charge undergoes renormalization group (RG) flow as envisaged by Zamolodchikov ( 1986), the monopole charge also undergoes RG flow. This induces a time variation of the non-quantized monopole charge which eventually leads to the anomalous velocity giving rise to the spin Hall effect. This is the situation which occurs when the vortex line is tilted with respect to the wave front propagation direction.

In sec. II we recapitulate certain features of the skyrmion model of a fermion and discuss its implications in the generation of relativistic electron vortex beams. In sec. III we consider the spin-orbit coupling and in sec IV we study the situation for tilted vortex lines with respect to the wave front propagation direction when the beam carries OAM in an arbitrary direction.

\vskip .5 cm

\section{Skyrmion model of a fermion, The geometric phase and relativistic electron vortex beam} \label{sec2}

\noindent
It has been shown in some earlier papers ( Bandyopadhyay and Hajra 1987, Hajra and Bandyopadhyay 1991) that in
Nelson's stochastic quantization procedure (Nelson  1966, 1967) the quantization of a fermion can be achieved when we introduce an internal variable that appears as a direction vector. This direction vector essentially gives rise to the spin degree of freedom. In fact this gives rise to the SL(2,C) gauge theory and demanding Hermiticity we may take the gauge field belonging to the unitary group SU(2). In this scenario spin degrees of freedom are represented as SU(2)
gauge bundle. This effectively represents a gauge theoretic extension of the space-time coordinate as well as momentum which can be written as gauge covariant operator acting on functions in phase space
\begin{equation}\label{eq1}
Q_\mu =  -i\left( \frac{\partial}{\partial p_\mu}+{\cal{A}}_\mu (p)\right), ~~~~~~
P_\mu =  i\left( \frac{\partial}{\partial q_\mu}+{\cal{B}}_\mu (q)\right)
\end{equation}
where ${\cal{A}}_\mu ({\cal{B}}_\mu)$ are the momentum(spatial coordinate)dependent SU(2) gauge field.  Here $q_\mu(p_\mu)$ denotes the mean position (momentum) of the external observable space. In this formalism a massive fermion appears as a skyrmion (Skyrme 1961, 1962). It is noted that the configuration variables as well as the momentum variables given by eqn. (\ref{eq1}) represent non-commutative geometry and the non-commutativity parameter is given by
\begin{equation}\label{eq2}
[Q_{\mu}, Q_{\nu}] =  {\cal{F}}_{\mu \nu}(p),~~~~
[P_{\mu}, P_{\nu}] =  {\cal{F}}_{\mu \nu}(q)
\end{equation}
where ${\cal{F}}_{\mu \nu}$ is the corresponding field strength given by ${\cal{F}}_{\mu \nu}=\partial_\mu {\cal {A}}_\nu- \partial_\nu{\cal {A}}_\mu +[{\cal {A}}_\mu, {\cal {A}}_\nu].$
The functional dependence of this non-commutativity parameter corresponds to the existence of monopoles (Jackiw 1985, Berard and Mohrbach 2004). In particular, the spatial components of the momentum variable can be taken to satisfy
\begin{equation}
[p_i,p_j] =i \mu \epsilon_{ijk}\frac{x_k}{r^3} \label{eq4}
\end{equation}
where $\mu$ represents the monopole strength.

The general property of a non-Abelian gauge theory is that when the topological $\theta$-term given by 
\begin{equation}\label{eq5}
L=\theta ^\star {\cal{F}}_{\mu \nu}{\cal{F}}_{\mu \nu},
\end{equation}
 where $*{\cal F}_{\mu \nu} =\frac{1}{2}\epsilon_{\mu\nu\lambda\sigma} {\cal F}^{\lambda \sigma}$ is the Hodge dual is introduced in the Lagrangian, this gives rise to a certain topologically nontrivial Abelian gauge field in the configuration space. The situation may be viewed as if in the gauge orbit space the position of  a particle indicated by  ${\cal{A}}$ (non-Abelian  gauge field) moves in the space $U$ of non-Abelian gauge potentials under the influence of an Abelian gauge potential. One can write for the large gauge transformation of an SU(2) gauge field ${\cal{A}}=g^{-1}dg+g^{-1}ag,$ where
${\cal{A}}={\cal{A}}_\mu d x^\mu$ and $a=a_\mu d x^\mu$ with $g(x)\in $SU(2). The gauge orbit space $U/G$ where $G$ denotes the space of local gauge transformations $g(x)$ consists of the points $a(x)$. This gauge orbit space is multiply connected and has the topology of a ring indicating that there is a vortex line which is topologically equivalent to a magnetic flux line (Wu and Zee 1985, Sen and Bandyopadhyay 1994). One may note that a gauge orbit space effectively represents a loop space, when a loop can be visualized as an orbit (Loll 1992). This suggests that when the topological $\theta-$ term is incorporated in the theory a vortex line equivalent to a magnetic flux line is enclosed by the loop. When a scalar particle encircles the loop enclosing the magnetic flux line, it acquires a geometric phase (Berry phase) apart from the usual dynamical phase. This geometric phase is given by $\phi_B=2\pi \mu,$ where $\mu$ is the monopole charge associated with the magnetic flux line and $\mu=1/2$ corresponds to one magnetic flux line (Banerjee and Bandyopadhyay 1992). So for $\mu=1/2$ the particle acquires the phase $e^{i\pi}$ which is the phase associated with a fermion when it undergoes a $2\pi$ rotation. Thus a fermion may be visualized as a scalar particle encircling a magnetic flux line. In this formalism a fermion represents a skyrmion.

In this skyrmionic picture of an electron, the electron vortex beam is a natural consequence. Indeed, it is possible to have a cone of plane waves with a vortex line having a fixed polar angle $\theta$ with the z axis. The scalar electron encircling the vortex line can carry quantized OAM. In fact the experimental observation of the electron vortex beam carrying OAM (Uchida and Tonomura 2010, Verbeeck et. al. 2010, McMorran et. al. 2011), substantiates the skyrmion model of an electron. Since the vortex line gives rise to the spin degrees of freedom, this essentially represents a spin vortex. As the characteristic feature of a fermion such as the sign change of the wave function after a $2\pi$ rotation is  a manifestation of the Berry phase acquired by the scalar electron orbiting around the vortex line, the fermionic properties do not change when the scalar electron carries quantized OAM. This follows the fact that total phase in this case is $2\pi l+ \phi_{B},$ where $l$ is the OAM and $\phi_{B}$ is the Berry phase acquired by the scalar particle when it traverses the loop around the vortex line. For quantized $l$ as $l\in Z,$ the factor $2\pi l$ gives rise to a trivial contribution and so the effective phase is essentially $\phi_{B}.$ So the electron vortex beam can carry quantized OAM without disturbing the fermionic feature. In this context, we may add that the vortex beams carrying large values of OAM have a very large region around the $z$ axis, where the wave function is effectively zero (McMorran, B. J et al , 2011). This gives rise to the situation when OAM carrying scalar electron appears to be detached from the spin degrees of freedom.

In our formulation, the Berry phase acquired by the scalar electron encircling the vortex line is $2\pi \mu,$ $\mu$ being the monopole charge. When the monopole is located at the origin of a unit sphere, the Berry phase is given by $\phi_{B} = \mu\Omega(C),$ where $\Omega(C)$ is the solid angle subtended by the closed contour at the origin which is given by \beq \Omega(C) = \int_{C}(1 - cos \theta)d\phi = 2\pi(1- cos\theta),\label{eq8}\eeq
where $\theta$ is the polar angle of the vortex line with the quantization axis(z axis). So for $\mu = \frac{1}{2},$ we have the phase
\beq \phi_{B} = \pi(1 - cos\theta)\label{eq9}.\eeq
This corresponds to the flux associated with the monopole passing through the surface spanning the closed contour. In fact we have
\beq \mu\Omega(C) = e\phi|_{\Sigma}\label{eq10},\eeq where $\phi|_{\Sigma}$ is the flux through the surface $\Sigma$ and can be written as
\beq \phi|_{\Sigma} = \int_{\Sigma} \vec{B} .d\vec{\Sigma}\label{eq11}.\eeq Transforming to a reference frame where the scalar electron is considered to be fixed and the vortex state (spin state) moves in the field of the magnetic monopole around a closed path, $\phi_{B}$ in equation (6) corresponds to the geometric phase acquired by the vortex state. The angle $\theta$ represents the deviation of the vortex line from the $z$ axis. Equating this phase $\phi_{B}$ in eqn (6) with $2\pi\mu$ which is the geometric phase acquired by the scalar electron moving around the vortex line in a closed path, we find that the effective monopole charge associated with a vortex line having polar angle $\theta$ with the $z$ axis is given by
\beq \mu = \frac{1}{2}(1 - cos\theta)\label{eq12}.\eeq

 In this connection, one may note that for a particle moving around a closed path, adiabaticity is not an essential criterion for acquiring the Berry phase. Indeed, Aharonov and Anandan (1987) have pointed out that the Berry phase is acquired even when the system is nonadiabatic in nature.   The relation (9) shows that for $\theta = 0$ and $\frac{\pi}{2}$ it takes quantized values but for other angles $0 < ~\theta<~\frac{\pi}{2}$ $\mu$ is non-quantized. We note that when the vortex line representing the spin axis is along the $z$-direction,  i.e when the vortex line is parallel to the wave propagation direction (implying $\theta = 0$) we have the paraxial vortex beam. For
$\theta = \frac{\pi}{2}$ the vortex line is orthogonal to the wave propagation direction. For other values of $\theta$ corresponding to non-quantized monopole charge the vortex line is tilted in an arbitrary direction. This implies the deviation of the spin axis from the $z$ axis and represents the anisotropic feature associated with the system. These three states correspond to the screw, edge and mixed edge-screw dislocations in optical beams.

\section{Relativistic electron vortex beam and spin orbit interaction}
A paraxial relativistic electron vortex beam carrying quantized OAM involves a scalar electron and the vortex line along the axis of propagation implying that the polar angle $\theta = 0.$ The corresponding beams essentially represent a superposition of monoenergetic plane waves. In this case OAM and SAM are separately conserved. However, we can introduce spin-orbit interaction(SOI) which will lead to non-paraxial vortex beams. Indeed, Bliokh et al.(2011) studied the situation by solving the Dirac equation using cylindrical coordinates. For convenience we here recapitulate some of the results derived by these authors. We consider the Dirac equation (c = $\hbar$ = 1)
\beq i\partial_{t}\psi = \left(\vec{\alpha} . \vec {p} + \beta m\right)\psi \label{13},\eeq where $\vec{\alpha}$ and $\beta$ are the $4\times 4$ Dirac matrices, $\vec{p} = -i\partial_{\vec{r}}$ is the momentum operator and $m$ is the electron mass. The positive energy eigenstates are \beq \psi_{\vec{p}}(\vec{r}, t) = W(\vec{p})exp\left[i(\vec{p}.\vec{r} - Et)\right]\label{14},\eeq with energy $E = \sqrt{p^{2} + m^{2}}$  and $ W = \frac{1}{\sqrt{2}} \left( \begin{array}{cr}
   \sqrt{1 + \frac{m}{E}}w \\
    \sqrt{1 - \frac{m}{E}}\vec{\sigma}.\vec{\kappa}w

   \end{array} \right).~~~~ \\ \label{15}$
Here $\vec{\sigma}$ are the Pauli matrices, $\vec{\kappa} = \frac{\vec{p}}{p}$ is the momentum direction vector and $w$ is the two-component spinor characterizing the electron polarization in the rest frame with $E = m.$ Using cylindrical coordinates $(r, \varphi, z)$ in real space and $(p_{\perp} , \phi, p_{\parallel}) = (p sin\theta, \phi, p cos\theta)$ in momentum space we write the Fourier spectrum
\beq \tilde{\psi}_{l}(\vec{p}) = \frac{1}{i^{l}p_{\perp 0}}W(\vec{p})\delta(p_{\perp} - p_{\perp 0})exp(il\phi)\label{16},\eeq
where $p_{\perp 0} = p sin\theta_{0},$ $p_{\parallel 0} = p cos\theta_{0},$ $\theta_{0}$ being the fixed polar angle with the $z$ axis. Beams represent superposition of plane waves which are uniformly distributed over the azimuthal angle $\phi \in (0, 2\pi)$ with a vortex phase dependence $exp(il\phi),$ $l = 0, \pm 1, \pm 2...$ The beam field is given by the Fourier integral of $\tilde{\psi}_{l}(\vec{p})$ \beq {\psi}_{l}(\vec{r}, t) = \frac{exp(i\Phi)}{2\pi i^{l}}\int_{0}^{2\pi}W(\vec{p})exp\left[i\xi cos(\phi - \varphi) + il\phi\right]d\phi\label{17},\eeq
where $\Phi = (p_{\parallel 0}z - Et)$ and $\xi = p_{\perp 0}r.$ Assuming that the polarization amplitudes are the same for all the plane waves we have
\beq \psi_{l} = \frac{exp i\Phi}{\sqrt 2}\left[\left(\begin{array}{cr}
   \sqrt{1 + \frac{m}{E}}w \\
    \sqrt{1 - \frac{m}{E}}\sigma_{z}cos\theta_{0}w

   \end{array}\right)e^{il\varphi}J_{l}(\xi) + i \left(\begin{array}{cccr}
     0  \\
    0 \\
 -\beta\sqrt{\Delta}\\
       0

\end{array}\right)e^{i(l -1)\varphi}J_{l-1}(\xi) +i \left(\begin{array}{cccr}
           0  \\
             0 \\
             0 \\
\alpha\sqrt{\Delta}
           \end{array}\right)e^{i(l +1)\varphi}J_{l+1}(\xi)\right] \label{18}\eeq
where $\Delta = (1-\frac{m}{E})sin^{2}\theta_{0}.$
The first term in the square bracket represents a scalar like Bessel beam of the order of $l$: $\psi_{l} \propto J_{l}(\xi)e^{i(l\varphi + \Phi)}. $ The terms proportional to $\sqrt{\Delta}$ describe the polarization dependent coupling  implying SOI.

In the present framework we note that when the configuration of an electron is viewed as a scalar particle encircling a vortex line in a specific direction, the solution of the Dirac equation in cylindrical coordinates exhibits the vortex-dependent properties explicitly. It is noted that when OAM $l$ is taken to be zero so that the SOI is switched off the solution (\ref{18}) reduces to the term in the relativistic limit $\frac{m}{E} \rightarrow 0$
\beq \psi = \frac{e^{i(pz - Et)}}{\sqrt{2}}\left(\begin{array}{cc}
            w \\
           \sigma_{z}w

           \end{array}\right),\eeq
where $w$ represents the electron polarization. As the vortex line gives rise to the spin degrees of freedom we note that the spinorial term involving $w$ corresponds to the contribution of the vortex line where the plane wave $e^{i(pz - Et)}$ represents the contribution of the scalar electron orbiting around the vortex line which is taken to be along the $z$ axis. When the polar angle of the vortex line with the $z$ axis  $\theta \neq 0$ and the scalar electron orbiting around it carries any arbitrary OAM with $l\in Z,$ the first term in the square bracket in (\ref{18}) in the relativistic limit $\frac{m}{E} \rightarrow 0$ reduces to
\beq \psi_{l} = \frac{exp i\Phi}{\sqrt 2}e^{il\varphi}\left(\begin{array}{cr}
   w \\
   \sigma_{z}cos\theta w

   \end{array}\right)J_{l}(\xi).\label{200}\eeq
Here the term involving $w$ corresponds to the contribution of the vortex line and the factor $\frac{exp i(\Phi + l\varphi)}{\sqrt 2}J_{l}(\xi)$ corresponds to the contribution of the scalar electron. The second and third term in the square bracket in (\ref{18}) corresponds to the contribution of the SOI.

It is noted that the SOI factor $\Delta$ is determined by the Berry phase. In fact in the relativistic limit $\frac{m}{E}\rightarrow 0$ we find $\Delta = sin^{2}\theta = 4\mu(\mu -1)$ which follows from eqn (9). Here $\mu$ is the monopole charge associated with the Berry phase. For $\mu = 0(1),$ corresponding to $\theta = 0(\pi),$ we have $\Delta = 0$ which indicates that the vortex line is parallel (anti-parallel) to the $z$ axis and SOI vanishes. For $\mu = \frac{1}{2},$ corresponding to $\theta = \frac{\pi}{2}$ the vortex line is orthogonal to the $z$ axis. However $\Delta$ involves non-quantized monopole charge for other values of $\theta.$ It may be recalled  here that the association of SOI with the Berry phase has been studied by other authors in some earlier works. It has been pointed out that
when we treat the orbital degrees of freedom and the spin degrees of freedom as slowly and fast varying variables respectively, the effective Hamiltonian for the slow degrees of freedom may contain a nontrivial gauge potential which represents the Berry connection. The SOI is derived by making an adiabatic approximation to the Dirac equation for an electron moving in a smooth external potential in which the orbital degrees of freedom are treated as slowly varying with respect to the spin degrees of freedom (Mathur 1995). Bliokh et al. (2011), have described the OAM and SAM with Berry phase corrections and predicted SOI in relativistic electron vortex beams. The SOI term has been incorporated through the gauge theoretic methods by several authors (Fujita 2009, Obata 2008 and Tan 2008). This effectively leads to the interaction of the spin with a magnetic field in momentum space. In the present formalism the SOI is obtained through the Berry phase, which essentially corresponds to a monopole and thus manifests the presence of a magnetic field. In view of this fact we note that this formalism effectively has the same underlying physics as in the gauge theoretical formalism adopted by other authors.

 In the present formalism we study SOI from the gauge theoretic extension of the space coordinates as discussed in the previous section. In 3D space we can write from (\ref{eq1})
\beq
\vec{Q} = \vec{q} + {\cal \vec{A}}(\vec{p}),~~ {\cal \vec{A}}(\vec{p})\in SU(2)\label{20}.\eeq

The generalized OAM can now be written as
\beq \vec{\tilde{L}} = \vec{Q} \times \vec{p} = \vec{q}\times \vec{p} + {\cal \vec{A}}(\vec{p})\times \vec{p} = \vec{L} + \vec{L}^{'} \label{21},\eeq
with \beq \vec{L}^{~'} =  {\cal \vec{A}}(\vec{p})\times \vec{p}\label{22}.\eeq	
The gauge field ${\cal \vec{A}}(\vec{p})$ can be written as
\beq {\cal \vec{A}}(\vec{p}) = \vec{A}(\vec{p}) \times \vec{\sigma}\label{23},\eeq
with $\vec{A}(\vec{p}) = \mu \frac{\vec{p}}{p^{2}},$ $\mu$ being the monopole charge in the momentum space. From this we find \begin{equation}\label{24}
\vec{L}^{~'} = {\cal \vec{A}}(\vec{p})\times \vec{p}~~
= \vec{A}(\vec{p}) \times \vec{\sigma}\times \vec{p}~~
= \mu \frac{\vec{p}}{p^{2}}\times \vec{\sigma}\times \vec{p}
.\end{equation} 		
Now substituting $\frac{\vec{p}}{p} = \vec{\kappa},$ $\vec{\kappa}$ being the unit vector, we have
\beq \vec{L}^{~'} = -\mu \vec{\kappa}\times (\vec{\kappa}\times \vec{\sigma})\label{25}.\eeq
The expectation value of $\vec{\sigma}$ i.e $\langle\vec{\sigma}\rangle$ is given by
$ \langle\vec{\sigma}\rangle = \frac{\langle\psi|\vec{\sigma}|\psi\rangle}{\langle\psi|\psi\rangle}\label{26},$
where $\psi$ is a two-component spinor
$ \psi = \left(\begin{array}{cr}
   \psi_{1} \\
    \psi_{2}
 \end{array}\right) \label{27}$ with $\langle\psi|\psi\rangle = 1.$
 This gives \beq \langle\vec{\sigma}\rangle = \langle\psi|\vec{\sigma}|\psi\rangle = \vec{n}\label{28},\eeq
 with $\vec{n}^{2} = 1.$
 We can write from (\ref{25})
 \beq \langle\vec{L}^{'}\rangle = -\mu \vec{n}\label{29}.\eeq
 Now taking into account the mapping $(\hbar = 1 )$
 \beq \vec{L}\rightarrow l\hat{\vec{z}}, ~~~\vec{S}\rightarrow s\hat{\vec{z}} \label{30}\eeq and using the relation $\vec{\tilde{L}} + \vec{\tilde{S}} = \vec{L} + \vec{S}$ we find from equation (\ref{29}) and (\ref{30})
 \begin{equation}\label{100}
 \langle\vec{\tilde{L}}\rangle = (l - \mu)\hat{\vec{z}},~~~~~~~~~~~~~~~~
 \langle\vec{\tilde{S}}\rangle = (s + \mu)\hat{\vec{z}}
 \end{equation}
This suggests that a part of the angular momentum is transformed from the SAM to OAM implying SOI. It may be mentioned that the quantized value of $\mu = \frac{1}{2}$ corresponds to the relation $|\mu| = s.$ It can be noted that for non-quantized value of $\mu,$ the expectation value  $ \langle\vec{\tilde{L}}\rangle$ and $\langle\vec{\tilde{S}}\rangle$ can take arbitrary values.

For the quantized values of $\mu,$ the expression for $\langle \vec{\tilde{L}}\rangle$ expresses the relation for the angular momentum in the presence of a magnetic monopole. Indeed when the OAM $l = 0,$ the quantized value for $\mu = \frac{1}{2}$ suggests the total angular momentum $\frac{1}{2}$ indicating that the intrinsic angular momentum of the system is $\frac{1}{2}$ which is the SAM of an electron.

Essentially, when the scalar electron traverses a closed loop around the vortex line acquires the Berry phase as \beq \phi_{B} = \oint \vec{A}(\vec{p})d\vec{p} = 2\pi\mu \label{31}.\eeq

The expressions (\ref{100}) and (\ref{31}), explicitly exhibit that the Berry phase plays a significant role in SOI. 

From the Dirac eqn. the general features of SOI can be studied from the Foldy-Wouthuysen transformations (FW) momentum representation (Bliokh 2005, Berard and Mohrbach 2006) separating the positive and negative energy components. This effectively leads to the noncommutativity of space when the spatial coordinates are given by $\vec{R} = \vec{r} + \vec{A},$ with $\vec{A} = \frac{\vec{p}\times \vec{\sigma}}{2p^{2}}(1-\frac{m}{E}),$ where $\vec{A}$ is the non-Abelian Berry connection but not a monopole. However, $\vec{A}$ is here a momentum dependent gauge field which incorporates spin degrees of freedom through the relation $\vec{S} = \frac{1}{2}\vec{\sigma},$ $\vec{S}$ being the spin operator and $\vec{\sigma}$ is the vector of Pauli matrices. The Berry phase is obtained by integrating $\vec{A}$ along the contour of the Bessel beam spectrum in momentum space and is found to be $\phi_{B} = \int \vec{A} . d\vec{p} = 2\pi\Delta s$ (Bliokh 2011). Here $\Delta s$ is the spin variable which modifies the OAM, so that the expectation $\langle\vec{L}\rangle$ is now given by $\langle\vec{L}\rangle = (l + \Delta s)\hat{\vec{z}},$ with $\langle L_{z}\rangle = (l + \Delta s).$ In our formulation the Berry connection corresponds to a monopole and the monopole charge is linked with the helicity. In fact, the angular momentum of the charged particle in the field of a magnetic monopole is given by $\vec{J} = \vec{L} - \mu\hat{\vec{r}},$ where $\vec{L}$ is the OAM and $\mu$ is the monopole charge. When OAM is zero, the total angular momentum of the system which is effectively the spin degrees of freedom of the system is given by $|\mu|$ with $S_{z} = \pm \mu.$ In view of this, the Berry phase obtained in terms of the spin variable $\Delta s$ in Dirac equation formalism can be taken to correspond to the phase obtained in terms of the monopole charge $\mu.$ Thus we observe that the Berry connection derived in Dirac equation formalism essentially leads to the same physics as envisaged in the present formalism.  

\section{Tilted vortex and spin Hall effect}

We have elucidate in sec II, that when the polar angle $\theta$ of the vortex line with the $z$-axis is not $0$ and $\frac{\pi}{2},$ the Berry phase involves non-quantized value of the monopole charge, which we denote as $\tilde{\mu}$. In earlier papers (Bandyopadhyay 2000, 2011) it has been pointed out that the monopole charge in $3+1$ dimensions is equivalent to the central charge $c$ of the conformal field theory in $1+1$ dimensions. Zamolodchikov(1986) has shown that the central charge $c$ undergoes the renormalization group (RG) flow. In analogy to this, it has been pointed out that the monopole charge $\mu$ undergoes an RG flow. This suggests that when $\mu$ depends on a certain parameter $\lambda$ we have

\noindent- $\mu$ is stationary at fixed points $\lambda^{*}$ of the RG flow.\\
- at the fixed points $\mu(\lambda^{*})$ is equal to the monopole charge $\mu$ given by the quantized values $0$, $\pm\frac{1}{2},$ $\pm 1$....\\
-$\mu$ decreases along the RG flow i.e $L\frac{\partial \mu}{\partial L}\leq 0,$ where $L$ is a length scale. \\
Now transforming the length scale to the time scale $(L = ct)$ we can write
\beq t\frac{\partial\mu}{\partial t}\leq 0\label{32},\eeq which implies that we can consider $\mu$ as the time dependent parameter. In fact when $\mu$ takes a value on the RG flow and is non-quantized (denoted by $\tilde{\mu}$), we can take it as a function of time $\tilde{\mu}(t)$ and at certain fixed value of time it takes the quantized value $\mu.$

It is now clear that the explicit time dependence of the monopole charge, $\tilde{\mu}(t)$ effectively makes the corresponding gauge field explicitly time dependent. An electric field $\vec{{\cal E}}$  is now generated through its derivative $\frac{\partial\vec{A}}{\partial t}.$
An electric field accelerates electrons so that the momentum carries explicit time dependence.
We denote the time dependent momentum as $\vec{k}.$ In this case we can introduce a non-inertial coordinate frame with basis vectors $(\vec{v}, \vec{w}, \vec{t})$ attached to the local direction of momentum $\vec{t} = \frac{\vec{k}}{k}.$ This coordinate frame rotates as $\vec{k}$ varies with time. Such rotation with respect to a motionless (laboratory) coordinate frame describes a precession of the triad $(\vec{v}, \vec{w}, \vec{t}),$ with some angular velocity. Now taking the direction of the vortex line at an instant of time as the local $z$ axis which represents the direction of propagation of the wave front, we note that this corresponds to the paraxial beam in the local frame. In this local non-inertial frame the local monopole charge will correspond to a pseudospin. Indeed, the expectation value of the spin operator
\beq \left\langle \vec{S}\right\rangle = \frac{1}{2}\frac{\langle\psi|\vec{\sigma}|\psi\rangle}{\langle\psi|\psi\rangle}\label{43},\eeq
undergoes precession with the precession of the coordinate frame. This suggests that the polarization state depends on the choice of the coordinate frame (Bliokh 2009). When the direction of the vortex line is taken to be the local $z$ axis in the non-inertial frame, the local value of $\tilde{\mu}$ is changed to the quantized value $|\mu| = \frac{1}{2}$ due to the precession of the spin vector and thus corresponds to the pseudospin  in this frame. The pseudospin vector $\vec{S}$ is parallel to the momentum vector $\vec{k}.$ Now transforming the momentum $\vec{p}$ by the time dependent momentum $\vec{k}$ in the gauge potential, we can write the Berry curvature as
\beq \vec{\Omega}(\vec{k}) = \mu\frac{\vec{k}}{k^{3}}.\label{40}\eeq In fact the time dependence of $\tilde{\mu}$ is here incorporated through the time dependence of $\vec{k}.$ This curvature will give rise to an anomalous velocity given by
\beq \vec{v}_{a} = \dot{\vec{k}}\times \vec{\Omega}(\vec{k}). \label{41}\eeq
This anomalous velocity gives rise to the spin Hall effect. Indeed substituting the expression of $\vec{\Omega}(\vec{k})$ given by eqn. (\ref{40})in (\ref{41}), we can write the anomalous velocity as
\beq \vec{v}_{a} = \mu \dot{\vec{k}}\times\frac{\vec{k}}{k^{3}} . \label{42}\eeq
 Thus the anomalous velocity is perpendicular to the pseudospin vector and points along opposite directions depending on the chirality  $s_{z} = \pm \frac{1}{2}$ corresponding to $\mu> 0(< 0).$ This separation of the spins gives rise to the spin Hall effect (Dyakonov and Perel 1971). Thus a tilted vortex line with respect to the propagation direction in the inertial frame carrying OAM will give rise to relativistic spin Hall effect. This essentially describes the polarization dependent shift of the wave trajectory.

The SOI in the non-relativistic limit can be derived from the Dirac equation by introducing Foldy-Wouthuysen (FW) transformation (Foldy and Wouthuysen 1950) separating the positive and negative energy components in the Dirac equation. The transformation
 $ \psi^{'} = U_{FW}(\vec{p})\psi$
 with  $ U_{FW} = \frac{1}{\sqrt{2}}\left(\sqrt{1 +\frac{m}{E}} - \beta\vec{\alpha}.\vec{\kappa}\sqrt{1 -\frac{m}{E}}\right)\label{139}$ diagonalizes the Dirac Hamiltonian
$ U^{\dagger}_{FW}\left(\vec{\alpha.\vec{p}} + \beta m\right)U_{FW} = \beta E\label{140}$ and also yields $ W^{~'} = U^{\dagger}_{FW}W = (w, 0)^{T}\label{141}$ for the plane wave (11). Using the projection on the positive energy subspace which excludes the negative energy levels, the electron position operator is given by (Bliokh 2005, Berard and Mohrbach 2006)
 \beq \vec{R} = \vec{r} + {\cal{\vec{A}}}(\vec{p})\label{142}\eeq
 with \beq {\cal{\vec{A}}}(\vec{p}) = \frac{\vec{p}\times\vec{\sigma}}{2p^{2}}(1 - \frac{m}{E})\label{143}.\eeq
 The non-relativistic spin operator is given by $\vec{S} = \frac{1}{2}\vec{\sigma}$ and the SOI is introduced through the operator
 \beq \vec{\Delta} = {\cal{\vec{A}}}(\vec{p})\times\vec{p} = - (1 - \frac{m}{E})\vec{\kappa}\times(\vec{\kappa}\times\vec{S})\label{144}.\eeq
The spin dependent connection ${\cal{\vec{A}}}(\vec{p})$ gives rise to the Berry phase when the integration is performed along the contour of the Bessel beam spectrum in momentum space (Bliokh 2011) \beq \phi_{B} = \oint{\cal{\vec{A}}}(\vec{p}) d\vec{p} = 2\pi\Delta s \label{145}.\eeq
We can now have simple mapping
\beq \vec{L}\rightarrow l\hat{\vec{z}},~~ \vec{S} \rightarrow s \hat{\vec{z}},~~ \vec{\Delta} \rightarrow \Delta s \hat{\vec{z}}\label{146}.\eeq
The SOI is now determined by the spin to orbital angular momentum conversion analogous to that depicted in eqn (26) where $\mu$ is replaced by $-~\Delta s.$

In our framework, when the relativistic electron is depicted as a scalar particle moving around a vortex line and the space-time coordinate is given by a gauge theoretic extension, where the corresponding gauge field ${\cal A}(\vec{p})\in SU(2)$, the non-relativistic effect is obtained in the sharp point limit. In the sharp point limit the residual effect of the non-Abelian gauge field is manifested through an Abelian gauge field (Bandyopadhyay 1990). The Abelian gauge field in the momentum space gives rise to a momentum dependent magnetic field. The SOI, which is derived in the non-relativistic limit through the momentum dependent gauge field obtained through the FW transformation can now be incorporated through the momentum dependent magnetic field. The non-relativistic Hamiltonian (for detailed derivation see the appendix) in presence of the SOI term can now be written as (Fujita at el. 2011, Basu. B and Chowdhury. D 2013, Chowdhury. D and Basu. B, 2013)
\beq H = \frac{p^{2}}{2m} + V  - g\vec{\sigma}.\vec{B}(\vec{p})\label{147},\eeq where $g$ is the coupling constant and $V$ is the electrostatic potential. Time dependence of the momentum will make the magnetic field $\vec{B}(\vec{p})$ time dependent. We can have the explicit time dependence of the momentum when we switch over to the interaction picture. We can split the Hamiltonian in $H$ into two parts $H_{0} + H_{1},$ so that for $H_{0}$ we take \beq H_{0} = e\vec{{\cal E}}.\vec{r}\label{34}\eeq and  \beq H_{1} = \frac{\vec{p}^{2}}{2m} - g\vec{\sigma}.\vec{B}(\vec{p}). \label{35}\eeq Here $\vec{{\cal E}}$ is the electric field. In the interaction picture an operator $O$ in the Schrodinger picture is transformed as
\beq \tilde{O}(t) = e^{iH_{0}t} O e^{-iH_{0}t} \label{36}.\eeq
It is noted that the state vector $|\psi(t)\rangle$ in the Schrodinger picture is now transformed as
\beq |\tilde{\psi}(t)\rangle = e^{iH_{0}t}|\psi(t)\rangle \label{38}.\eeq
For the momentum operator it gives
\beq \vec{k}(t) = \vec{p} - e\vec{{\cal E}}t.\label{37}\eeq

We can now rewrite the expression for the anomalous velocity generated by the Berry curvature $\vec{\Omega}(\vec{k})$ given by eqn (31) involving explicitly
the non-relativistic spin vector $\vec{S}$ and the electric field $\vec{\cal{E}}.$ In fact from eqn (43) we have,
\beq \dot{\vec{k}} = -e\vec{\cal{E}}\label{53}.\eeq
The monopole charge $\mu$ effectively represents the electron helicity as this corresponds to $s_{z} = \pm\frac{1}{2}$ in the local frame for $\mu> 0$($\mu<$ 0) we can substitute $\mu$ in eqn (31) by the expression of the helicity
\beq \mu = \frac{\vec{S}.\vec{k}}{2|k|}. \label{54}\eeq
Inserting this in eqn (31)the expression of the anomalous velocity yields
\beq \vec{v}_{a} = \frac{\vec{S}\times e\vec{\cal{E}}}{2k^{2}}.\label{55}\eeq
This gives rise to the spin current along the direction of $\vec{v}_{a}$ when two opposite orientations of the spin move in opposite directions.

The expression of $\vec{v}_{a}$ in (31) can be rewritten in terms of the unit vector $\vec{t} = \frac{\vec{k}}{|\vec{k}|}$ and its time derivative $\dot{\vec{t}}$ as (Bliokh 2009)
\beq \vec{v}_{a} = \mu\frac{\vec{\dot{k}}\times\vec{k}}{k^{3}} = \mu \dot{\vec{t}}\times\vec{t}  .\label{56}\eeq Denoting $\frac{\dot{\vec{t}}}{|\dot{\vec{t}}|} = \vec{n},$ we note that the spin current is orthogonal to the local plane ($\vec{t}, \vec{n}$). Thus we observe that for a tilted vortex with respect to the wave propagation direction though the Berry phase acquired by the orbiting scalar electron may be viewed as an artefact of a rotating coordinate frame, the spin Hall effect is a Coriolis type transverse deflection as the spin current is orthogonal to the local plane $(\vec{t},\vec{n})$. Furthermore, the Berry phase involves non-quantized monopole charge when the vortex line remains non-rotating in an inertial frame. But in a rotating frame one can take the quantized value $|\mu| = \frac{1}{2}$ by choosing the proper coordinate frame when the vortex line is taken to be in the propagation direction. However the spin Hall effect represents a real deflection of the wave trajectory as the spin current moves out of plane in the orthogonal direction of the local $(\vec{t},\vec{n})$ plane and thus becomes independent of the local coordinate frame.

In addition, One may indicate here that a well defined gauge field in time space which has the physical significance of an effective magnetic field accounts for the SHE in the Rashba system in the adiabatic limit (Fujita et al 2010). This magnetic field is also found to be the underlying origin of the anomalous velocity due to the curvature in momentum space. In fact the anomalous velocity due to the Berry curvature in momentum space is a direct manifestation of the effective magnetic field from the gauge field in the time space. Thus in our analysis, the time dependence of the monopole charge,  which gives rise to SHE through the generation of anomalous velocity from the Berry curvature, effectively reveals to an analogous situation as developed through the introduction of a gauge field in time space (Fujita et al 2010).
\section{Discussion}
We have studied here the geometrodynamics of the relativistic electron vortex beam from the perspective of the geometric phase associated with the scalar electron encircling the vortex line. It is observed that the vortex beam is a natural consequence of the skyrmion model of an electron as investigated in the quantization scheme of a fermion. In this scenario a fermion is visualized as a scalar particle encircling a vortex line which is topologically equivalent to a magnetic flux line. The geometric phase acquired by the scalar particle moving around the vortex line in a closed loop essentially gives rise to the specific properties of a fermion such as the sign change of the wave function after a $2\pi$ rotation as well as the spin-statistics relation. The electron vortex beam can carry OAM as the introduction of the quantized OAM given by $l\in Z$ contributes only trivially to the total phase so that the effective phase is the concerned geometric phase. This ensures the fermionic properties of such a system. It is pointed out that we have paraxial beam when the vortex line is parallel to the wave front propagation direction. In this case both the OAM and the SAM are separately conserved. However when the spin-orbit coupling is switched on we have non-paraxial beam. When the vortex line is orthogonal to the propagation direction the monopole related to the Berry phase is quantized. Interestingly, when the vortex line is tilted with respect to the propagation direction of the wave front, the related Berry phase involves non-quantized monopole. From an analysis of the RG flow of the monopole charge, it is argued that in this case we have temporal variation of the direction of the vortex line giving rise to spin Hall effect which can be observed experimentally.

However, these features associated with electron vortex beams are analogous to optical vortex beams. Indeed, the paraxial beam corresponds to the screw wave front dislocation and the non-paraxial beam associated with the quantized monopole charge represents edge dislocation when the vortex line is orthogonal to the wave propagation direction. The tilted vortex lines with respect to the wave propagation direction which involve non-quantized monopole charge correspond to the mixed edge-screw dislocations in optical beams. These incorporate temporal variations. It has been pointed out by Bliokh and Nori(2012), that one can avoid temporal diffraction, when these dislocations represent spatio-temporal vortex beams which can be achieved through the Lorentz transformation of the corresponding spatial beams. However for mathematical consistency one has to take into account the "relativistic Hall effect" caused by a shift in the transverse direction of the OAM carrying object observed in a moving frame. We have argued above that in case of electron vortex beams with tilted vortices the temporal variation leads to spin Hall effect.

Finally, one can conclude that the dynamics of relativistic electron vortex beam is determined by the Berry phase acquired by the scalar electron orbiting around the vortex line. When this phase involves quantized monopole we have paraxial and non-paraxial beam such that in the former(latter) case the vortex line is parallel  (orthogonal) to the propagation direction. On the otherhand, when this phase involves non-quantized monopole we have tilted vortex lines which have temporal variation and causes the spin Hall effect. This non-quantized monopole is realized through the RG flow of the monopole charge and hence does not envisage the contribution of the Dirac string.
\begin{center}
References
\end{center}
Aharonov, Y and Anandan, J, 1987, Phys Rev Lett. {\bf 58}, 1593.\\
Allen. L,  Beijersbergen M. W,  Spreeuw R. J. C, and  Woerdman J. P
1992 Phys. Rev. A {\bf 45}, 8185.\\
Bandyopadhyay P, 1996 {\it Geometry, Topology and quantization}, Kluwer Academic, Dordrecht, The Netherlands\\
Bandyopadhyay P, 2000 Int. J. Mod Phys. A {\bf 15}, 1415--1434.\\
Bandyopadhyay P, 2010 Proc. R. Soc.(London) A. {\bf 466} 2917--2932.\\
Bandyopadhyay P, 2011 Proc. R. Soc.(London) A. {\bf 467}, 427--438 .\\
Bandyopadhyay, P. , Hajra, K. 1987 J. Math. Phys. {\bf 28}, 711–-716. \\
Banerjee, D. , Bandyopadhyay, P. 1992 J. Math. Phys. {\bf 33}, 990–-997. \\
Basu, B. Chowdhury, D 2013 Annals of phys {\bf 335}, 47–-60.\\
Berard, A. , Mohrbach, H. 2004 Phys. Rev. D {\bf 69}, 12701–-12704.\\
Berard, A. , Mohrbach, H. 2006 Phys. Lett. A {\bf 352}, 190.\\
Berry, M. V. 1984 Proc. R. Soc. Lond. A
{\bf 392}, 45–-57.\\
Bliokh.  K. Y. 2005 Europhys. Lett. {\bf 72}, 7.\\
Bliokh.  K. Y. 2009 J. Opt. A: Pure Appl. Opt. {\bf 11}, 094009. \\
Bliokh.  K. Y. et. al. 2007  Phys.Rev.Lett.{\bf 99}, 190404.\\
Bliokh.  K. Y. et. al. 2011 Phys. Rev. Lett. {\bf 107}, 174802.\\
Bliokh.  K. Y. and Nori. F 2012 Phys. Rev. A 86, 033824\\
Chowdhury, D and Basu, B, 2013  Annals of Physics {\bf 329}, 166.\\
Dyakonov. M. I and Perel. V. I, 1971 Sov. Phys. JETP Lett. {\bf 13}, 467. \\
Foldy, L.L, and Wouthuysen, S.A, 1950 Phys. Rev. D {\bf 78}, 29\\
Fujita. T,  Jalil. M B A and Tan S G, Murakami. S 2011 J. Appl. phys. {\bf 110}, 121301.\\
Fujita,T, 2009, J. Phys. Soc. Jpn vol. {\bf 78}, 104714.\\ 
Fujita,T, Jalil, M.B, and Tan, S.G, 2010 New J. Phys. {\bf 12}, 013016.\\
Hajra, K. , Bandyopadhyay, P. 1991 Phys. Lett. A {\bf 155}, 7–-14.\\
Jackiw, R. 1985 Phys. Rev. Lett. {\bf 54}, 159–-162.\\
Loll, R. 1992 {\it New loop approach to Yang Mills theory}. In {\it Proc. XIX Int. Colloq. on Group
Theoretical Methods in Physics}, vol. II (eds. A. del Olmo, M. Santader,  J. Mateos-Guilarte),
pp. 122 –- 129. Annales de Fisica, Monografias, Salamanca, Spain.\\
Mathur, H. 1991 Phys. Rev. Lett.{\bf 67}, 3325.\\
McMorran. B. J et. al 2011 science {\bf 331}, 192.\\
Nelson, E. 1966  Phys. Rev. {\bf 150}, 1079–-1085.\\
Nelson, E. 1967 {\it Dynamical theory of Brownian motion} Princeton University Press,Princeton, N.J.\\
Nye. J. F and Berry . M. V 1974 Proc. R. Soc.(London) A. {\bf 336}, 165.\\
Obata, K,2008, Phys. Rev. B, {\bf 77}, 214429.\\ 
Sen, K. , Bandyopadhyay, P. 1994  J. Math. Phys. {\bf 35}, 2270–-2281.\\
 Skyrme. T. H. R 1961 Proc. R. Soc.(London) A. {\bf 260}, 127--138.\\
Skyrme. T. H. R 1962 Nucl. Phys {\bf 31}, 556.\\
Tan,S. G, Jalil, M. B. A., Liu, Xiong-Jun, and Fujita, T, 2008, Phys. Rev. B vol. {\bf 78}, 245321.\\
Uchida. M and Tonomura A 2010 Nature {\bf 464}, 737.\\
Verbeeck. J, Tian. H and Schattschneider. P, 2010 Nature {\bf 467}, 301.\\
Wu. Y.S and Zee. A, 1985  Nucl. Phys. B {\bf 258}, 157-168.\\
Zamolodchikov. A 1986 JETP Lett.{\bf 43}, 731.\\
\appendix
\section{Foldy Wouthusen transformation}
The  Dirac Hamiltonian for a particle of charge $e$ is given by

\begin{eqnarray}
H &=& \beta mc^{2} +  c\left(\vec{\alpha}.\vec{p}\right)
 + eV(\vec{r}) \label{q},
\end{eqnarray}
 where $ \beta$,  $\alpha $ are usual  Dirac matrices and $\Sigma$ is the  spin operator for 4-spinor.
To obtain a Hamiltonian in the low energy limit, one has to apply a series of Foldy-Wouthuysen  Transformations  (FWT) on the Hamiltonian (\ref{q} ).

The Dirac wave function is a four component spinor with the up and
down spin electron and hole components. Generically the energy gap between the
electron and hole is much larger than the energy scales associated with
condensed matter systems. One
can achieve this by block diagonalization method of the Dirac Hamiltonian
exploiting FWT.
The Hamiltonian (\ref{q} ) can be divided into  block diagonal and off
diagonal
parts denoted by $ \epsilon $ and $ O $, respectively,
\begin{equation}
H = \beta mc^{2} + O +\epsilon ,~~~
 O = c\vec{\alpha}.\vec{p}~~~
\epsilon = eV(\vec{r}).
\label{fwh}
\end{equation}
where $\beta=\gamma_0$
and $\alpha_i=\gamma_0\gamma_i$. Applying FWT on $H$ yields,
\begin{equation}
H_{FW} = \beta \left(mc^{2}+\frac{O^{2}}{2mc^{2}}\right)+ \epsilon
-\frac{1}{8m^{2}c^{4}} \left[O ,[O,\epsilon]\right]
\end{equation}

Calculating various terms of the Hamiltonian ($ O^{2},$ $\left[O,\epsilon\right],$ and $\left[O,[O,\epsilon]\right]$ ), following relation $ (\vec{\alpha} . \vec{A})(\vec{\alpha} .\vec{B}) =  \vec{A}.\vec{ B} +
i\vec{\Sigma}.(\vec{A}\times\vec{B} )$ we can write the
FW transformed Hamiltonian upto $\frac{1}{c^{2}}$ terms as
\begin{equation}
H_{FW} = \beta\left( mc^{2} + \frac{\vec{p}^{2}}{2m}\right) + V(\vec{r})
-  \frac{\hbar^{2}}{8m^{2}c^{2}}
(\vec{\nabla}.\vec{E}) -
\frac{i\hbar^{2}}{8m^{2}c^{2}}\vec{\Sigma}.(\vec{\nabla}\times \vec{E})-
\frac{\hbar}{4m^{2}c^{2}}\vec{\Sigma}.(\vec{E}\times \vec{p})\label{w}
\end{equation}
Consideration of constant electric field can help us leaving  the terms with  $(\vec{\nabla}\times \vec{E})$ and
$(\vec{\nabla}.\vec{E})$. Finally, we land up with the Hamiltonian for the upper component of Dirac spinor as
\begin{equation}
H_{FW} = \left( mc^{2} + \frac{\vec{p}^{2}}{2m}\right) + V(\vec{r})-
\frac{\hbar}{4m^{2}c^{2}}\vec{\sigma}.(\vec{E}\times \vec{p})\label{w1}
\end{equation}
This FW transformed
Hamiltonian gives  the dynamics of an electron (or hole with proper sign of
$e$) in the positive energy part of the full energy spectrum. Here $\vec{\sigma}$ is the Pauli spin matrix. 

For electrons moving through the lattice, the electric field $\vec{E}$ is Lorentz transformed to an effective magnetic field $(\vec{p}\times \vec{E})\approx \vec{B}(\vec{p})$ in the rest frame of electron. Thus from (\ref{w1}) we can write (Fujita at el. 2011, Basu. B and Chowdhury. D 2013, Chowdhury. D and Basu. B, 2013) the Hamiltonian as
\beq H = \frac{\vec{p}^{2}}{2m} + V(\vec{r})-
g\vec{\sigma}. \vec{B}(\vec{p}),\eeq
where the rest energy term is neglected and g is the coupling strength. 
\end{document}